\documentclass[aps,pre,twocolumn,groupedaddress]{revtex4-1}

\usepackage{graphicx}
\usepackage{bm}
\usepackage{amssymb}

\usepackage{color} 

\begin{document}


\title{Complete Delocalization in a Defective Periodic Structure}


\author{Behrooz Yousefzadeh}
\email[]{behroozy@caltech.edu}
\author{Chiara Daraio}
\email[]{daraio@caltech.edu}
\affiliation{Division of Engineering and Applied Science, California Institute of Technology, Pasadena, California, 91125 USA}


\date{\today}

\begin{abstract}

We report on the existence of stable, completely delocalized response regimes in a nonlinear defective periodic structure. In this state of complete delocalization, despite the presence of the defect, the system exhibits in-phase oscillation of all units with the same amplitude. This elimination of defect-borne localization may occur in both the free and forced responses of the system. In the absence of external driving, the localized defect mode becomes completely delocalized at a certain energy level. In the case of a damped-driven system, complete delocalization may be realized if the driving amplitude is beyond a certain threshold. We demonstrate this phenomenon numerically in a linear periodic structure with one and two defective units possessing a nonlinear restoring force. We derive closed-form analytical expressions for the onset of complete delocalization and discuss the necessary conditions for its occurrence. 

\end{abstract}



\maketitle

\section{Introduction}
\label{sec:intro}

The dynamic response of periodic structures (lattices) is prone to localization phenomena through different scenarios. For linear systems, localization occurs {\color{black}most readily} as a result of breaking the periodicity of the system, in defective or disordered lattices. In a defective system, {\color{black}at least one unit cell of the lattice is different from the others in its inertial or elastic properties} and the response localizes to the defective unit~\cite{defectMontroll}. In a disordered system, {\color{black}all unit cells deviate randomly from the reference unit cell} and localization occurs in an ensemble-average sense~\cite{andersonRev,anderson50,anderson}. For perfectly symmetric lattices, localization may still occur if the lattice is nonlinear, resulting in time-periodic solutions of the system that are localized in space -- these are called discrete breathers (DB) or intrinsic localized modes (ILM). This nonlinear localization phenomena may occur in both free and forced (externally driven) lattices, as well as infinite or finite lattices~\cite{todayDB,DBcolloq,DBchaos}.  
{\color{black}
Even in a linear lattice, existence of spatially localized states have been reported in perfectly symmetric systems~\cite{flatExp1,flatExp2}. These localized states are generated  by manipulating the geometry of the unit cell and correspond to non-dispersive portions of the dispersion relation known as flat bands; see~\cite{flatSimple} for more details. 
}

The focus of this work is on the dynamics of one-dimensional defective lattices. A key characteristic of defective lattices is the spatial localization of response amplitude to the defective unit~\cite{defectMontroll}. This localization corresponds to existence of a spatially localized mode shape with a natural frequency that lies outside the phonon spectrum (pass band) of the system. 
Defective lattices have been the subject of various investigations due to their interesting nonlinear wave dynamics. Examples include, among many others, stability and bifurcation analysis of nonlinear defect modes~\cite{defectBifurcation}, interaction of solitons with defects in photonics~\cite{defectSolitonPhotonics,solitonRevMalomed}, breather modes in defective granular chains~\cite{defectGranularDB,defectGranularJob,defectGranularNick}, nonlinear wave characteristics in defective systems~\cite{masonDefect}, non-destructive defect identification in granular media~\cite{defectNDT}, mechanical systems with tunable stiffness~\cite{chiaraExtreme}, exploiting the supratransmission phenomenon (\cite{supra_00,JSV1}) in a defective system to construct an all-mechanical switch~\cite{nickNature}, and autonomous magnetomechanical frequency converters~\cite{airhockey}. A dynamic response with a spatially localized profile is the centerpiece in all these examples. 

{\color{black}
In the present paper, we demonstrate that defect-borne localization may be completely eliminated by careful placement of a nonlinear element within the periodic structure. The ensuing delocalized state is stable and has a spatially uniform profile with all units moving in phase with each other. We refer to this as the state of complete delocalization. 
}

As the first step, to introduce the concept of complete delocalization and the key ingredients, we consider the simplest defective periodic system that can exhibit this phenomenon. This is a linear periodic system with a single nonlinear defect. The mathematical model of this system is presented in Section~\ref{sec:model}, along with the localization norm 
used in this study. 
We present the complete delocalization phenomenon in Section~\ref{sec:numerics} using numerical results. In Section~\ref{sec:analysis}, we develop closed-form analytical expression for predicting the onset of complete delocalization and obtain the required necessary conditions in terms of system parameters. We show in Section~\ref{sec:double} that complete delocalization may occur in systems with more than one defects. While the majority of the paper deals with damped-driven systems, we show in Section~\ref{sec:free} that complete delocalization may also be observed in the free response of the system. 
We discuss generalizations and summarize our findings in Sections~\ref{sec:discussion} and~\ref{sec:conclusion}.

\section{Mathematical Model}
\label{sec:model}

We consider a one-dimensional periodic structure (lattice) that is subject to uniform, harmonic excitation at all units. Experimentally, this can be achieved, for example, via excitation of the base in mechanical structures~\cite{EnglishCuevas} or optically driving a
lattice of charges by a harmonic electric field~\cite{page}. We further assume that the periodic structure is lightly damped and has a finite length ($N$ units). Although the assumption of finite length is not a necessary ingredient for realizing complete delocalization, it makes the results more readily applicable. Except for the defective unit, all units are assumed to be linear and identical. The defect is represented by a nonlinear restoring force, which includes a non-zero linear component. The non-dimensional equations of motion for this periodic system can therefore be written as follows
\begin{equation}
	\label{eq:EOM1}
	\ddot{u}_n + 2 \zeta \dot{u}_n + k_n u_n + k_c \Delta(u_n) + \alpha_n u_n^3 = F \cos(\Omega t)
\end{equation}
where $u_n(t)$ represents the displacement of the $n$-th unit for $1 \le n \le N$ and overdot denotes differentiation with respect to non-dimentional time $t$. Energy dissipation is modeled as uniform viscous damping in each unit with coefficient $\zeta$. Adjacent units are coupled linearly such that $\Delta(u_n)=2u_n-u_{n+1}-u_{n-1}$ for all units except at the boundaries where $\Delta(u_1)=u_1-u_2$ and $\Delta(u_N)=u_N-u_{N-1}$ (free boundary conditions). $k_c$ represents the strength of coupling between adjacent units. The external driving force is harmonic, with $F$ and $\Omega$ as its amplitude and frequency. The remaining stiffness parameters are defined as 
\begin{eqnarray*}
	 k_n         = 1 + \delta_{nn_d} k_d,  \\
	 \alpha_n = \delta_{nn_d} \alpha_d, 
\end{eqnarray*}
where $\delta_{nn_d}$ is the Kronecker delta and the defective unit is located at $n=n_d$. 
{\color{black}$k_d$ and $\alpha_d$ characterize the deviations of the elastic restoring force within the defective unit from the restoring force in other units. For ease of reference, we attribute this deviation to an internal force $F_d$ acting within the defective unit that we call the {defect force}. This defect force is a function of the motion of the defective unit only ($u_{n_d}$) and can be written as} 
\begin{equation}
	\label{eq:Fd}
	F_d=F_d(u_{n_d})=k_d u_{n_d} + \alpha_d u_{n_d}^3  .
\end{equation}
We have chosen the simplest possible form for $F_d$ that would allow complete delocalization. {\color{black}The defective unit is a Duffing oscillator~\cite{duffingBook} that is embedded within a linear lattice.}

We have chosen the following parameters for the periodic system in Eq.~(\ref{eq:EOM1}) throughout the paper: $\zeta=0.005$ (light damping), $k_c=0.01$ (weak coupling regime), $N=14$ (finite length). While operating near the anti-continuum limit (small $k_c$) is not necessary, it helps capture the dispersion effects of periodicity more easily for such a short periodic system -- also note that realizing this value for $k_c$ is practical~\cite{DBcolloq,EnglishCuevas,JSV1}. 
The remaining parameters are free.

There are various measures for quantifying the degree of localization. We use the inverse participation ratio~\cite{andersonRev}, IPR, which is a scalar defined by 
\begin{eqnarray*}
	\label{ipr}
	I\!P\!R = \frac{\sum^N_{n=1} (v_n^2)^2}{\left(\sum^N_{n=1} v_n^2\right)^2} \, , \quad \frac{1}{N} \le I\!P\!R \le 1
\end{eqnarray*}
where $v_n$ are the amplitudes of motion -- often \emph{IPR} is used for quantifying mode shape localization, in which case $v_n$ would represent an eigenvector of the periodic system. 
A value of $1/N$ denotes a uniform state where all units move with the same amplitude (complete delocalization) and a value of $1$ denotes the extreme localization state where all but one unit is motionless. 
Based on a rescaling of \emph{IPR}, we define an alternative measure
\begin{equation}
	M = -\log_N(I\!P\!R), \quad 0 \le M \le 1
\end{equation}
where $M=1$ occurs in the case of complete delocalization (uniform response) and $M=0$ occurs when only one unit is moving (complete localization).

In the absence of defect ($k_d = 0,\,\alpha_d=0$), the resulting dynamics has a spatially uniform state and $M=1$ at any forcing frequency $\Omega$; i.e. all units exhibiting harmonic, in-phase motion with identical amplitudes. In a linear defective system ($k_d \ne 0,\,\alpha_d=0$), $M<1$ at all driving frequencies $\Omega$. In a nonlinear defective system ($k_d \ne 0,\,\alpha_d \ne0$), we will show that if $k_d$ and $\alpha_d$ satisfy a certain condition, there exists a threshold on $F$ above which it is possible to obtain complete delocalization ($M=1$) at certain values of $\Omega$.

In this work, we denote the amplitude of vibrations for the $n$-th unit by $U_n$ and define it as 
\begin{eqnarray}
	\label{eq:Un}
	|U_n| = \sqrt{ \frac{2}{T} \int_0^T u_n^2(t) dt }
\end{eqnarray}
where $T=2\pi/\Omega$ is the period of vibrations. 
The response of the system in Eq.~(\ref{eq:EOM1}) can be computed numerically as a family of periodic orbits via pseudo-arclength continuation technique. {\color{black}Note that these solutions need not be harmonic and no restriction is imposed by the numerical procedure on the response other than periodicity.} We have used the software package AUTO to perform these computations. See~\cite{continuation_redBook,doedel} for further technical details.

\section{Delocalization in a lattice with a single defect}
\label{sec:numerics}

\begin{figure}[bt]
	\includegraphics[width=\linewidth]{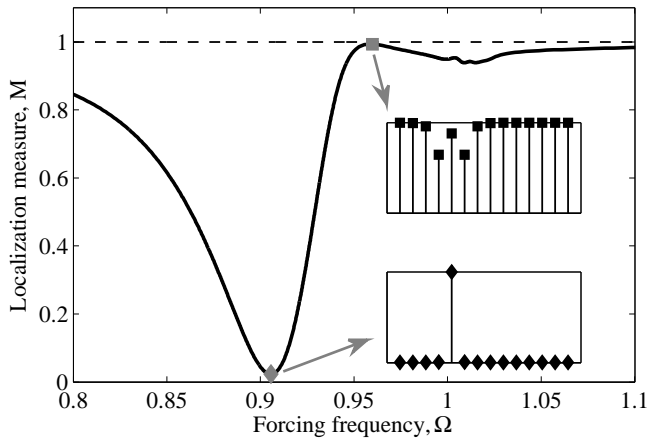}
	\caption{
		\label{fig1} 
		The localization measure $M$ for a (linear) system with $(n_d,k_d,\alpha_d)=(5,-0.2,0)$. The insets show the amplitude profiles corresponding to the maximum and minimum values of $M$, denoted by $\blacksquare$ and $\blacklozenge$ respectively. The amplitude profiles are normalized to have a maximum value of 1. 
	}
\end{figure}

\begin{figure}[bt]
	\includegraphics[width=\linewidth]{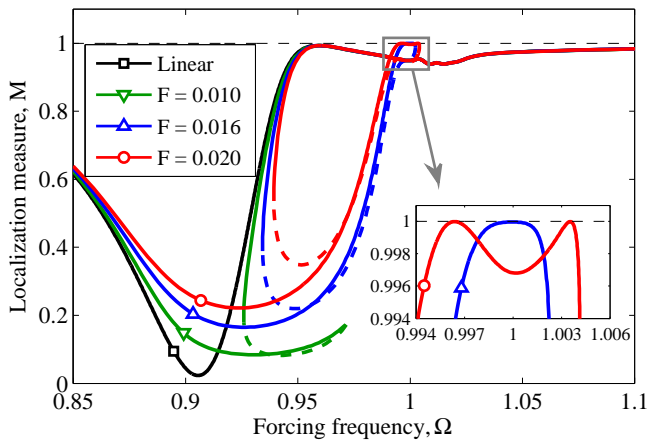}
	\caption{
		\label{fig2} 
		The localization measure for a (nonlinear) system with $(n_d,k_d,\alpha_d)=(5,-0.2,0.1)$ for increasing values of forcing amplitude $F$. The solid portions of each curve denote stable response and dashed portions denote unstable response. The linear response is repeated from Figure~\ref{fig1} for comparison. The inset zooms on the region where the state of complete delocalization is obtained at $M=1$. 
	}
\end{figure}

\begin{figure}[bt]
	\includegraphics[width=\linewidth]{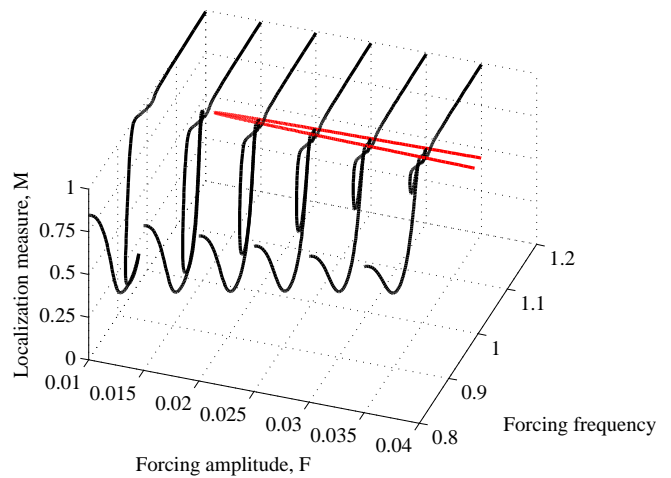}
	\caption{
		\label{fig3} 
		Evolution of the localization measure $M(\Omega)$ for the same system as in Figure~\ref{fig2} as a function of $F$. As the forcing amplitude increases beyond a threshold ($F>F^{\star}=0.016$), complete delocalization is achieved at two separate forcing frequencies. The red curve traces the intersection of $M(\Omega)$ with the $M=1$ plane, which is shown separately in Figure~\ref{fig4}. Stability information is not included in this graph. 
	}
\end{figure}

\begin{figure}[bt]
	\includegraphics[width=\linewidth]{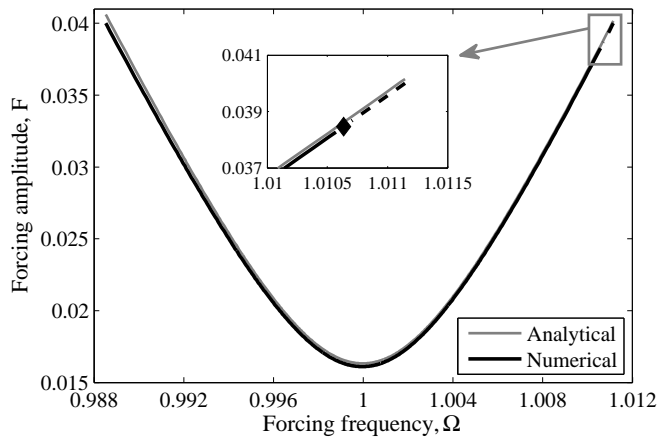}
	\caption{
		\label{fig4} 
		The threshold curve for the same system as in Figure~\ref{fig2}, showing the  forcing amplitude threshold for the onset of complete delocalization as a function of forcing frequency. The grey curve is obtained from Eq.~(\ref{eq:Fstar}). The black curve is obtained numerically with $M=1$ fixed. The solid portions denote stable response and dashed portions denote unstable response. The onset of instability is denoted by a diamond (cannot be captured by the analytical solution). 
	}
\end{figure}

\begin{figure}[bt]
	\includegraphics[width=\linewidth]{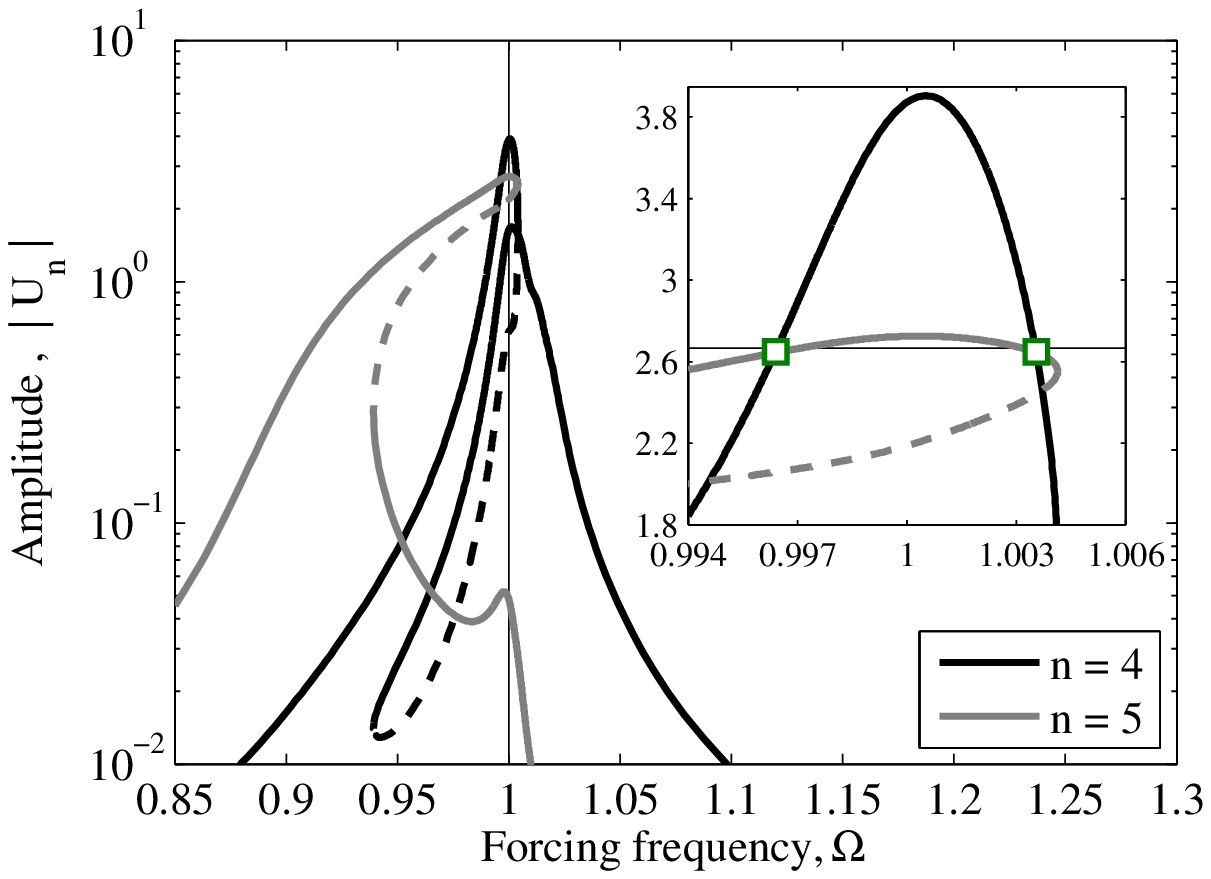}
	\caption{
		\label{fig5} 
		Steady state response amplitude of the same system as in Figure~\ref{fig2} at $F=0.020$, above the onset of complete delocalization. The stable portions of solution are shown using solid curves and dashed portions denote unstable response. The defect is located at $n_d=5$. 
		The inset zooms on the region where complete delocalization occurs {\color{black}(denoted by empty green squares)}, which correspond to $|U_5|=|U_4|$. The horizontal black line corresponds to the analytical prediction of response amplitude at complete delocalization by Eq.~(\ref{eq:U0star}). 
		}
\end{figure}

We start by considering a linear defective system with $k_d=-0.2$ and $\alpha_d=0$. 
Figure~\ref{fig1} shows the localization measure $M(\Omega)$ for this system for $0.80 \le \Omega \le 1.10$. The insets show the amplitude profiles corresponding to the maximum and minimum values of $M$, both occurring in this frequency range. The amplitude profiles are normalized to vary between 0 and 1.  
Maximum localization occurs at $\Omega = 0.9056$ with $M=0.0231$, with almost all the energy localized to the defective unit. {\color{black}This localization can also be understood as the resonance of the defective unit.} Minimum localization occurs at $\Omega=0.9598$. Despite the high value of $M=0.9936$ at this frequency, the amplitude profile is not uniform; neither do units oscillate in phase with each other. 

Figure~\ref{fig2} shows the localization measure for increasing values of forcing amplitude $F$ for a nonlinear defective system with $k_d=-0.2$ and $\alpha_d=0.1$. As the forcing amplitude increases, the amplitudes of motion increase (most importantly within the defective unit) and the nonlinear defect force becomes significant. This results in a bend in the $M(\Omega)$ curve towards higher frequencies due to the hardening nature of nonlinearity ($\alpha_d>0$). At $F=0.010$, we observe the influence of nonlinearity as appearance of additional solutions (bistability) and increase in the minimum value of $M$ in this range compared with the linear solution. Notice that there is little change in $M$ where the linear solution had a maximum near $\Omega=0.9598$. 
The additional branch of solutions grows with increasing $F$ and the localization measure approaches $M=1$ near $\Omega=1$. We can see this in Figure~\ref{fig2}, where complete delocalization is achieved near $\Omega \approx 1$ for $F=0.016$. 
Beyond the delocalization threshold, shown for $F=0.020$, there are two forcing frequencies at which complete delocalization can be achieved, one below and one above $\Omega=1$. For a given forcing frequency, though, we found only one state of complete delocalization.

Figure~\ref{fig3} shows $M(\Omega)$ curves for different values of $F$ for the same system parameters as in Figure~\ref{fig2}. For $F \ge 0.016$, these curves intersect the $M=1$ plane, tracing the locus of forcing amplitudes required for the onset of complete delocalization as a function of forcing frequency. 
This locus can be obtained as a two-parameter continuation in $F$ and $\Omega$ with $M=1$ fixed. This threshold curve lies on the $F-\Omega$ plane and is replotted in Figure~\ref{fig4}, including the stability information. We show in Section~\ref{sec:analysis} that the threshold curve is (approximately) a parabola in the $F-\Omega^2$ plane, with its vertex located slightly below $\Omega=1$. 

Figure~\ref{fig4} shows that the response of the system at the state of complete delocalization can become unstable (onset of instability denoted by a diamond). This instability occurs if the forcing amplitude is increased beyond a certain limit. 
For a fixed set of system parameters, we found the onset of instability to depend on the location of the defective unit. Most notably, this instability is triggered more easily (i.e. at a lower forcing amplitude) if the defect is located closer to the boundary of the lattice. 
The mechanism of instability is a torus (Neimark-Sacker) bifurcation, where a pair of complex conjugate Floquet multipliers exit the unit circle into the complex plane. Further analysis of the bifurcation structure of the response near this point is beyond the scope of our current investigation.

Figure~\ref{fig5} shows the steady state response amplitude at $n=5$  (defective unit) and $n=4$. We can see that the response of the defective unit has bent toward higher frequencies due to the hardening nature of the nonlinear force ($\alpha_d>0$) and, eventually, we have reached a point where $|U_5|=|U_4|$ near $\Omega = 1$. Interestingly, this is the linear natural frequency of the system when $k_d=0=\alpha_d$, implying that the defect force $F_d$ is zero at the state of complete delocalization. 
We will show in Section~\ref{sec:analysis} that this assumption can predict the onset of complete delocalization with very good accuracy. 

We also note the bistable nature of the response at the state of complete delocalization in Figure~\ref{fig5}. Complete localization corresponds to the higher-energy state of the system, which can be reached readily (i.e. without explicit knowledge of its basin of attraction) by up-sweeping the forcing frequency.

\section{Analytical Prediction of the Onset of Complete Delocalization}
\label{sec:analysis}

We can predict the onset of complete delocalization based on the nature of the response at $M=1$. In the linear system, complete delocalization can only be obtained in a perfectly periodic system. All units vibrate with equal amplitudes and phases, and the solution can be written as
\begin{equation}
	\label{eq:uniform}
	u_n(t) = U_0 \exp\left( i \Omega t \right)
\end{equation}
where $U_0$ is the uniform (complex-valued) amplitude of motion. Substituting this solution into the governing equations with $k_d=0=\alpha_d$, we obtain the following 
\begin{equation}
	\label{eq:U0}
	U_0 = \frac{F}{1-\Omega^2 + 2i \zeta \Omega}
\end{equation}
which is the familiar steady-state response of a single linear oscillator. Without loss of generality, we can assume that $F$ is real-valued. 
It is important to note that the coupling force vanished in the state of complete delocalization because adjacent units move in phase with each other; in other words, the coupling springs are not engaged at all. 

In the case of a linear defect (Figure~\ref{fig1}), as expected, neither the phase nor the amplitude of motion are equal among different units. Upon introduction of the nonlinear defect force, however, it becomes possible to achieve the state of complete delocalization again. This occurs because the defect force may vanish beyond a certain force threshold. 
Given that the response of the system is described by Eq.~(\ref{eq:uniform}) in this state, we can obtain the conditions for complete delocalization by substituting Eq.~(\ref{eq:uniform}) into Eq.~(\ref{eq:Fd}) and setting $F_d=0$. 
\begin{eqnarray*}
	F_d \approx (k_d  + 3/4 \alpha_d |U_0|^2 ) U_0\exp(i\Omega t)=0
\end{eqnarray*}
which results in 
\begin{eqnarray}
	\label{eq:U0star}
	|U_0^\star| = \sqrt{\frac{-4k_d}{3\alpha_d}}
\end{eqnarray}
where $|U_0^\star|$ denotes the amplitude of motion at the onset of complete delocalization. Note that the harmonic approximation used in the above analysis is valid for small (but finite) amplitudes of motion, and that the higher harmonics generated by the cubic nonlinearity have been neglected (the rotating wave approximation). 

We can immediately see from Eq.~(\ref{eq:U0star}) that a necessary condition for realizing complete delocalization is that 
\begin{eqnarray}
	\label{eq:condition}
	k_d \alpha_d<0
\end{eqnarray}
Based on Eq.~(\ref{eq:U0star}), the onset of complete delocalization occurs at the same amplitude of motion for all forcing frequencies. One would therefore expect all the points along the threshold curve in Figure~\ref{fig4} to have the same amplitude predicted by Eq.~(\ref{eq:U0star}). We found that the value of $|U_0^\star|$ based on numerical computations of Figure~\ref{fig4} (averaged over all units and forcing frequencies) is 1.2\% lower than the predicted value in Eq.~(\ref{eq:U0star}). We attribute this discrepancy mainly to the harmonic approximation used in the analysis and, to a lesser extent, to the fact that we used $M=0.99999$ for numerical computation of the threshold curve. 

Given that the defect force is zero in the state of complete delocalization, the expression in Eq.~(\ref{eq:U0}) remains valid for $|U_0^\star|$. We can therefore use Eqs.~(\ref{eq:U0}) and~(\ref{eq:U0star}) to obtain an expression for $F^\star$, the force threshold at the onset of complete delocalization, as follows
\begin{eqnarray}
	\label{eq:Fstar}
	F^{\star} = |U_0^\star| 
		\sqrt{ \Omega^{\star 4} -2 (1-2\zeta^2) \Omega^{\star 2}+1 }
\end{eqnarray}
Eq.~(\ref{eq:Fstar}) describes the threshold curve in the $F-\Omega$ plane, as shown in Figure~\ref{fig4}. Of course, the location of the secondary instability points (torus bifurcations) shown in Figure~\ref{fig4} cannot be predicted using the present analysis. 

The minimum forcing amplitude at which complete delocalization occurs ($F_{\text{cr}}$) can be obtained from Eq.~(\ref{eq:Fstar}) as 
\begin{eqnarray}
	\label{eq:Fcr}
	F_{\text{cr}} = \min\left( F^{\star} \right) = 2 |U_0^\star| \zeta \sqrt{1-\zeta^2 }
\end{eqnarray}
The corresponding critical forcing frequency is 
\begin{eqnarray}
	\label{eq:Fcr}
	\Omega_{\text{cr}} = \sqrt{1-2\zeta^2 }
\end{eqnarray}
The point $(F_{cr},\Omega_{cr})$ is the vertex of the threshold curve in Figure~\ref{fig4}.

We note that the location of the defect ($n_d$) and the length of the periodic system ($N$) do not play a role in the analysis above. This can in part be justified by recalling that, at the state of complete delocalization, the system has a uniform in-phase response across all units such that the coupling force vanishes. In this state, thus, adjacent units are not coupled to each other. We have indeed verified this numerically by computing the threshold curves for $1 \le n_d \le 7$ and calculating the relative standard deviation between these curves (standard deviation divided by the mean value) as a function of $\Omega$. We found the relative standard deviation to be smaller than 0.03\% for the left portion ($\Omega<1$) of the threshold curves and increasing to 0.28\% in the right portion ($\Omega>1$), where higher-order nonlinear effects appear. This confirms that, to a first approximation, the location of the defect has a negligible effect on the threshold curve. It is worth recalling that we found the main influence of $n_d$ and $N$ in triggering secondary instabilities, as already discussed in Figure~\ref{fig4}.

\section{Delocalization in a Lattice with Two Defects}
\label{sec:double}

\begin{figure*}[bt]
	\includegraphics[width=\linewidth]{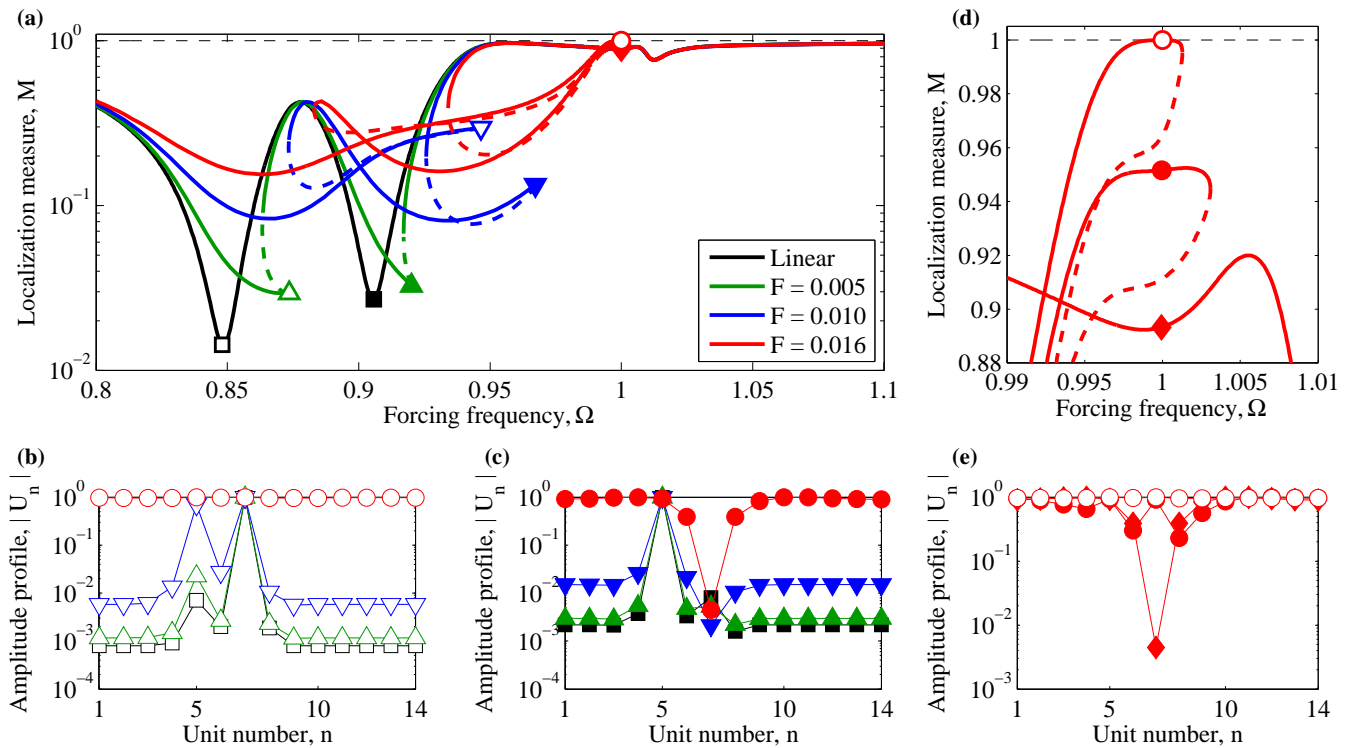}
	\caption{
		\label{fig:double} 
		Complete delocalization in a lattice with two defects, $n_{d1}=5$, $n_{d2}=7$. 
		\textbf{(a)}~$M(\Omega)$ for increasing values of $F$;  
		\textbf{(b)}~evolution of the amplitude profiles for the $n_{d2}$-localized mode;
		\textbf{(c)}~evolution of the amplitude profiles for the $n_{d1}$-localized mode; 
		\textbf{(d)}~$M(\Omega)$ at $F=0.016$ near the region where complete delocalization occurs; 
		\textbf{(e)}~amplitude profiles for the three stable solutions at complete delocalization. 
		All amplitude profiles are normalized to have a maximum value of 1. The markers in (b), (c) and (e) correspond to points in (a) and (d). 
	}
\end{figure*}

Complete delocalization may also be realized in a lattice with multiple defects, provided that the conditions derived in Section~\ref{sec:analysis} are satisfied. 
{\color{black}We show this in a system with two defective units. We consider the defects to have the same nature as the one in Section~\ref{sec:model}: deviations in their elastic properties with respect to that of other units. To characterize these deviations, we attribute them to internal forces acting within the defective units that we call defect forces. We consider the defect forces to have the same functional form as in Eq.~({\ref{eq:Fd}}) and denote them by $F_{d1}$ and $F_{d2}$, acting respectively on units $n_{d1}$ and $n_{d2}$. 
} 

In the state of complete delocalization, we expect the periodic system to have a uniform, in-phase amplitude profile with $M=1$. Following the same procedure as in Section~\ref{sec:analysis}, we arrive at the same expressions as in Eq.~(\ref{eq:U0star}) for the amplitudes of motion at which the two defect forces vanish. Equating these amplitudes, we arrive at the following condition for complete delocalization: 
\begin{eqnarray}
	\label{eq:double}
	\frac{k_{d1}}{\alpha_{d1}}=\frac{k_{d2}}{\alpha_{d2}}
\end{eqnarray}
To demonstrate complete delocalization in a lattice with two defects, we take $(n_{d1},k_{d1},\alpha_{d1})=(5,-0.2,0.1)$, identical to the defect used in Section~\ref{sec:numerics}. The second defect is described by $(n_{d2},k_{d2},\alpha_{d2})=(7,-0.3,0.15)$, in accordance with Eq.~(\ref{eq:double}). 
{\color{black}For clarity of demonstrations, we chose $k_{d2}$ such that the two defect frequencies are visibly separate and lie below the pass band (see Figure~\ref{fig:double}(a)). Although there is no restriction on the relative signs of $k_{d1}$ and $k_{d2}$, realizing $k_{d1}k_{d2}<0$ experimentally might be unnecessarily cumbersome, if not impractical. Also, the main effect of $n_{d2}$ is in determining the stability of the response, similar to what we observed for a single defect.}
All other system parameters are the same as those used in Section~\ref{sec:numerics}. 

Figure~\ref{fig:double}(a) shows the evolution of $M$ as a function of forcing parameters $\Omega$ and $F$. For the linear lattice {\color{black}(black curve with square marker)}, there are two main localized states, each corresponding to localization to one of the defect sites. The corresponding amplitude profiles are shown in Figures~\ref{fig:double}(b) and~\ref{fig:double}(c). As $F$ increases, the nonlinear forces become significant, resulting in increase of the defect natural frequencies ($\alpha_{d1}\!>\!0,\alpha_{d2}\!>\!0$). This manifests in Figure~\ref{fig:double}(a) as bending of the response curve at locations corresponding to the two defects {\color{black}for $F=0.005$ (green curve with up-triangle marker) and $F=0.010$ (blue curve with down-triangle marker}. The amplitude profiles corresponding to these points are shown in Figures~\ref{fig:double}(b) and~\ref{fig:double}(c), where we can notice the decrease in the degree of localization as $F$ increases. As we reach $F^{\star} \approx 0.016$, the branch corresponding to the $n_{d1}$-localized mode reaches $M=1$ near $\Omega \approx 1$ and the state of complete delocalization is achieved {\color{black}(red empty circles in Figures~\ref{fig:double}(a) and~\ref{fig:double}(d))}.

Figure~\ref{fig:double}(d) zooms on the $M(\Omega)$ curve at $F=0.016$ near $\Omega=1$, where complete delocalization occurs. Notice that the system has three stable periodic solutions (limit cycles) in this region. Figure~\ref{fig:double}(e) shows the amplitude profiles of the delocalized state along with the other two stable solution at the same frequency $\Omega^{\star}$. Only one of these three stable solutions corresponds to a delocalized state; similar to the single-defect case, the higher-energy state is delocalized (cf. Figure~\ref{fig5}). 

We note that there is no particular link between response tristability and complete delocalization. Indeed, we can see regions of tristability in Figure~\ref{fig:double}(a) for both $F=0.010$ and $F=0.016$. A tristable region occurs due to energy dependence of response, in the same spirit that the commonly-known nonlinear bistability occurs. 
The occurrence of tristability in driven periodic systems was previously shown theoretically and experimentally in a chain of coupled pendula~\cite{tristability}.

The threshold curve for the onset of complete delocalization (the locus of $M=1$ in the $F-\Omega$ plane) was computed in the same way as described in Section~\ref{sec:numerics}. The resulting threshold curve is similar to Figure~\ref{fig4} and is not shown again. We found two major differences between the threshold curves of the single-defect and double-defect systems: (a) the frequency at which instabilities occur along the right half of the threshold curve ($\Omega>1$), though the instability mechanism remains the same; (b) the threshold curve for the double-defect system terminated at $\Omega \approx 1.01$ near a torus (Neimark-Sacker) bifurcation point. We were not able to find a state of complete delocalization at higher frequencies. 
A detailed analysis of the bifurcation structure near this point lies beyond the framework of our current study.

\section{Delocalization of the Free Response}
\label{sec:free}

\begin{figure}[bt]
	\includegraphics[width=\linewidth]{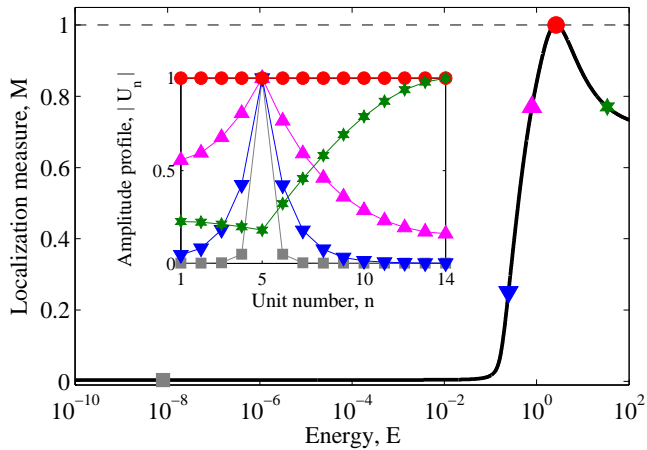}
	\caption{
		\label{fig:free} 
		Evolution of the defect mode as a function of energy. The inset shows the amplitude profiles corresponding to the markers in the main graph. All amplitude profiles are normalized to have a maximum value of 1. $M=0.25$ at the point with blue down-triangle marker, while both the magenta up-triangle and green hexagonal markers correspond to the same value of $M=0.77$. 
	}
\end{figure}

Complete delocalization may also be observed in the free response of an undamped system; i.e. $F=0$, $\zeta=0$ in Eq.(\ref{eq:EOM1}). At low energies (motions with small amplitudes), the system has $N$ normal modes with frequencies $\omega_n$. As the energy of the system increases, these modes evolve and the defect mode may reach the state of complete delocalization. 

In the absence of defect ($k_d=0$, $\alpha_d=0$), we have $1 \le \omega_n < \sqrt{1+4k_c}$ and the lowest mode $\omega_1=1$ is completely delocalized. Upon introduction of a linear defect ($k_d \ne 0$, $\alpha_d=0$), one of the modes becomes strongly localized to the defective unit, with its natural frequency lying outside the pass band (phonon spectrum). For $k_d=-0.2<0$ used in this study, we have $\omega_d=0.905<1$ and $M=0.004$ for the linear localized mode. For the nonlinear defective system ($k_d \ne 0$, $\alpha_d \ne 0$), the defect mode delocalizes as the amplitudes of motion increase until $M=1$ is reached at a specific energy level. Simultaneously, the defect frequency $\omega_d$ increases with energy and approaches 1, which is the first natural frequency of the defect-free system. 

Figure~\ref{fig:free} shows the evolution of the defect mode as a function of the total energy in the system, $E$, along with the amplitude profiles at different energy levels. We have defined $E$ as 
\begin{eqnarray*}
	E = \frac{1}{N} \sum_{n=1}^N |U_n|^2
\end{eqnarray*}
where $|U_n|$ is defined in Eq.~(\ref{eq:Un}). For small energies ($E<0.1$), the nonlinear force in the defective unit is negligible and the response of the system is very similar to that of the linear system (grey square marker). As energy increases more, the nonlinear defect mode gradually delocalizes until it reaches complete delocalization (uniform, in-phase vibration) at $E^{\star}=2.64$. 
The defect mode re-localizes beyond this point with a different amplitude profile (green hexagonal star). We have not investigated the evolution of the defect mode beyond the point of complete delocalization. 

It is worth noting that the amplitude at which complete delocalization occurs, and the corresponding energy value $E^{\star}$, agree very well with the analytical prediction in Eq.~(\ref{eq:U0star}). The difference between the analytical and numerical values of $|U_0^{\star}|$ is about 0.5\%. This is the amplitude at which the defect force $F_d$ vanishes.

We also computed the nonlinear evolution of defect modes in the double-defect system of Section~\ref{sec:double}. We found that the first defect mode (localized to $n_{d1}$) became completely localized at the predicted energy level $E^{\star}$, while the second defect mode (localized to $n_{d2}$) did not delocalize completely. The second mode reached the maximum value of $M=0.862$ at $E=4.02>E^{\star}$. These results are in  agreement with the findings of Section~\ref{sec:double}.

\section{Discussion}
\label{sec:discussion}

We presented the complete delocalization phenomenon in a very basic setting: a one-dimensional linear periodic system possessing a nonlinear defect with cubic restoring force -- see Eqs.~(\ref{eq:EOM1}--\ref{eq:Fd}). We found the principle ingredient for occurrence of complete delocalization to be a vanishing defect force. This condition is indeed not limited to the setting described by Eqs.~(\ref{eq:EOM1}--\ref{eq:Fd}). For example, if the main lattice (i.e. apart from $F_d$) is nonlinear, one would still expect that obtaining the state of complete delocalization is possible provided that the defect force is chosen carefully. There would be more possibility of nonlinear resonances in this scenario, and care should be taken to avoid them. Another generalization is to lattices with multiple defects. The case of two defects was presented in Section~\ref{sec:double}. The condition for complete delocalization in Eq.~(\ref{eq:double}) extends to lattices with multiple defects provided that the defect force has the same form as in Eq.~(\ref{eq:Fd}). 

We used free-free boundary conditions to ensure that the first mode of the defect-free lattice ($F_d=0$) is indeed described by Eq.~(\ref{eq:uniform}). This mode is completely delocalized and remains unchanged as $N$ varies; thus, it coincides exactly with the lower edge mode of the same lattice with infinite units. For other boundary conditions (e.g. fixed boundary conditions), the first mode of the finite lattice cannot be described by Eq.~(\ref{eq:uniform}) and a uniform response amplitude cannot be realized in the finite lattice even in the absence of defect. Although using a periodic boundary condition would also be feasible (a ring arrangement), its experimental realization is more involved than a free-free boundary condition; e.g. see~\cite{EnglishCuevas,JSV1} for straight-forward realizations of free boundary conditions in base-driven mechanical systems. A very good candidate for experimental realization of complete delocalization is a chain of coupled cantilevers with controllable magnet-induced nonlinear defect force. 

Although we studied complete delocalization in the weak coupling regime (small $k_c$), there are no restrictions on the strength of coupling. We verified this by computing the evolution of the defect mode in the same system as in Figure~\ref{fig:free} but with $k_c=0.10$: we obtained the same result. This happens because $\Delta(u_n)=0$ in Eq.~(\ref{eq:EOM1}) in the state of complete delocalization; thus, the coupling forces vanish. In this sense, the units become effectively decoupled. This also explains why the length of the lattice ($N$) and the location of the defect(s) within the lattice ($n_d$) do not change the threshold curves; they only influence the stability of delocalized solutions. 

The strength of coupling is important -- from a purely linear perspective -- in relation to the linear portion of the defect force, determined by $k_d$. The main parameter in this sense is the ratio $k_d/k_c$. For otherwise fixed system parameters, the influence of the defect force on the linear dynamics of the lattice decreases as $k_d/k_c$ is decreases. Notice that this has no phenomenological bearing on realizing the state of complete delocalization described here.

An important consideration in complete delocalization is to distinguish between hardening ($\alpha_d>0$) and softening ($\alpha_d<0$) types of nonlinearity. The obvious consideration is given by Eq.~(\ref{eq:condition}) that ensures the nonlinear component of the defect force can cancel its linear component. 
An inconspicuous consideration, on the other hand, comes from the relation of the defect frequency with respect to the pass band. For a defect with hardening nonlinearity, the defect frequency is below the pass band ($\omega_d<1$) and the instabilities in the threshold curve appear on the opposite side of the pass band (i.e. for $\Omega>1$) -- see Figure~\ref{fig4}. We made a similar observation in a system with a softening defect nonlinearity, $(n_d,k_d,\alpha_d)=(5,0.2,-0.1)$, where $\omega_d>\sqrt{1+4k_c}$. The numerically computed threshold curve for this system was the same as the one shown in Figure~\ref{fig4}, except for the stability information. In the softening system, the solution along the threshold curve (i.e. the completely delocalized solution) lost stability through a Neimark-Sacker bifurcation close to the upper edge of the pass band and remained unstable for all forcing frequencies below that.

The analysis of the nonlinear evolution of defect modes of the system, i.e. the free-response analysis, can provide useful insights into the complete delocalization phenomenon, as seen in Section~\ref{sec:free}. We found the analysis of free response to be more straight forward for finding completely delocalized states. From a practical point of view, however, capturing the nonlinear behavior of normal modes is not an easy task, at least in mechanical systems. The stability of completely delocalized solutions is also less problematic in driven systems due to presence of damping.

When analyzing the free response of the softening system, we were not able to reach complete delocalization in the system. The defect mode delocalized considerably as a function of energy, reaching a maximum value of $M \approx 0.85$ at $E \approx 1.79$. This is the value of $M$ for the $N$-th mode of the linear defect-free system. We note that the defect modes approach the closest linear mode of the defect-free system as $E$ increases. This is the lower edge mode for the hardening system and the upper edge mode in the softening system. Only the lower linear edge mode is completely delocalized in the defect-free system. Thus, we could only observe complete delocalization in the free response of the {\color{black}system with hardening nonlinearity}. This also explains why only the first mode of the double-defect system could delocalize completely.

{\color{black}

Up to here, our discussion of complete delocalization was exclusively focused on the first (in-phase) mode of the periodic system, described by Eq.~(\ref{eq:uniform}). This is because complete delocalization as described by $M=1$ can only be achieved by the in-phase mode in a free-free lattice. In order to make generalizations to other modes, we recall that complete delocalization is accompanied by vanishing of defect forces ($F_d=0$). With that in mind, we can view complete delocalization as \emph{retrieval of the linear, defect-free response within a nonlinear, defective lattice}. In this generalized perspective, the previous condition of $M=1$ becomes a special case of $M_{\text{nonlinear}}=M_{\text{linear}}$. An example of this generalization was given in the previous paragraph for the out-of-phase mode of a softening lattice where the linear defect-free modes of the system were reached at $E \approx 1.79$. 

In principle, it should be possible to extend this generalization to the case of a force-damped system. This would be achieved by modifying the external force such that it could excite other lattice modes: replacing the right-hand side of Eq.~(\ref{eq:EOM1}) with $f_n\cos(\Omega t)$, where $f_n$ is proportional to one of the mode shapes of the system. In practice, however, this is limited to exciting either the in-phase or out-of-phase mode. The analysis of generalized complete delocalization for the out-of-phase mode ($\pi$-mode) is also tractable analytically. In this case, we would replace Eq.~(\ref{eq:uniform}) by $u_n(t)=U_0(-1)^n\exp(i\Omega t)$ and follow the same procedure as in Section~\ref{sec:analysis}. Of course, one major difference here would be that the strength of coupling plays a very significant role. It is not clear to us if a similar analysis could be easily extended to the other modes of the system. 

Finally, we demonstrated the existence of stable, completely delocalized states in a defective lattice using numerical simulation and approximate analysis. Nevertheless, from a purely mathematical perspective, an exact analytical proof remains to be presented. 

}

\section{Conclusion}
\label{sec:conclusion}

We introduced the phenomenon of complete delocalization in a nonlinear defective lattice. {\color{black}This refers to existence of a stable response regime within a defective periodic system that is characterized by a spatially uniform amplitude and phase profile}. We showed that the spatially-localized response associated with the presence of a linear defect may be eliminated by careful placement of a nonlinear element within the defective unit. 
{\color{black}
Energy is uniformly distributed throughout the lattice in the ensuing delocalized state. 
}

Complete delocalization may be observed in both the free and forced responses of the periodic system. In damped-driven systems, the elimination of defect-borne localization may be realized provided that the driving amplitude is beyond a certain threshold. In the free response of the system, the defect modes become completely delocalized at a certain energy level. 

{\color{black}
We characterized the defects as internal forces acting within the defective units. We showed that complete delocalization occurs when these defect forces vanish. This allowed us to develop closed-form analytical expressions to predict the onset of complete delocalization and obtain necessary conditions for the phenomenon to occur. Our analytical results showed excellent agreement with numerical analysis of the system. We further generalized the concept of complete delocalization as retrieval of the linear defect-free response within a nonlinear defective system.
}

Experimental realization of the complete delocalization phenomenon is in progress. 
Complete delocalization {\color{black}(and its generalization)} opens new avenues for manipulating the propagation of mechanical waves in phononic crystals and mechanical metamaterials. 


\begin{acknowledgments}
B.Y. was supported by a postdoctoral fellowship from the National Science and Engineering Research Council of Canada. 
This material is based upon work supported by the National Science Foundation under EFRI Grant No.~1741565.
\end{acknowledgments}

\bibliography{J04_Idefect}

\end{document}